\documentclass[a4paper,11pt]{article}
\usepackage{fullpage,graphicx}

\newtheorem{theorem}{Theorem}
\newtheorem{corollary}[theorem]{Corollary}
\newtheorem{observation}[theorem]{Observation}

\begin{document}

\title{Ruler Wrapping}
\author{Travis Gagie, Mozhgan Saeidi and Allan Sapucaia}
\maketitle

\begin{abstract}
In 1985 Hopcroft, Joseph and Whitesides showed it is NP-complete to decide whether a carpenter's ruler with segments of given positive lengths can be folded into a line of at most a given length, such that the folded hinges alternate between 180 degrees clockwise and 180 degrees counter-clockwise.  At the open-problem session of 33rd Canadian Conference on Computational Geometry (CCCG '21), O'Rourke proposed a natural variation of this problem called {\em ruler wrapping}, in which all folded hinges must be folded the same way.  In this paper we show O'Rourke's variation has an linear-time solution.  We also show how, given a sequence of positive numbers, in linear time we can partition it into the maximum number of substrings whose totals are non-decreasing.
\end{abstract}

\section{Introduction}
\label{sec:introduction}

\begin{figure}[t]
\begin{center}
\includegraphics[width=\textwidth]{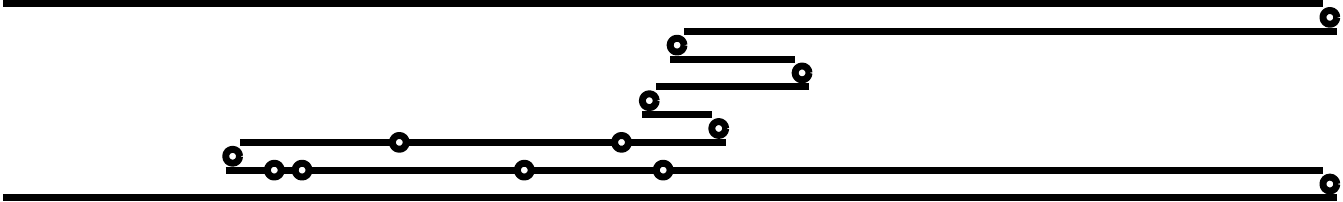}
\caption{A carpenter's ruler with segments of lengths 48, 24, 5, 6, 3, 4, 8, 6, 2, 1, 8, 5, 24 and 48, which can be folded into a line of length 96 because $\{5, 6, 3, 4, 8, 6, 2, 1, 8, 5\}$ can be partitioned into two subsets with equal sums.}
\label{fig:hardness}
\end{center}
\end{figure}

Problems about carpenters' rulers are a staple of computational geometry.  For example, in 1985 Hopcroft, Joseph and Whitesides~\cite{HJW85} posed the following question: can a carpenter's ruler whose segments have given positive lengths be {\em folded} into a line of at most a given length, with folded hinges alternating between 180 degrees clockwise and 180 degrees counter-clockwise (segments of the ruler having width 0 and folds being points)?  They showed this problem is NP-complete in the weak sense via a reduction from {\sc Partition}, as illustrated in Figure~\ref{fig:hardness}; gave a pseudo-polynomial algorithm for it; and gave a linear-time 2-approximation algorithm.  C\u{a}linescu and Dumitrescu~\cite{CD05} later gave an FPTAS for it.

At the open-problem session of the 33rd Canadian Conference on Computational Geometry (CCCG '21), Professor Joseph O'Rourke proposed a natural variation of this problem, in which all folded hinges must be folded the same way (either all 180 degrees clockwise or all 180 degrees counter-clockwise); he named this variation {\em ruler wrapping} and, as far as we know, it had not been considered before.

\section{A quadratic algorithm}
\label{sec:quadratic}

Suppose the ruler has $n$ segments and we lay it out flat, considering the hinges from left to right and pretending there is a hinge 0 at the left end and a hinge $n$ at the right end.  For each hinge, we define the {\em wrapping length} at that hinge to be the smallest length into which the segments to its left can be wrapped such that we can still fold the hinge.

\begin{figure}[t]
\begin{center}
\includegraphics[width=\textwidth]{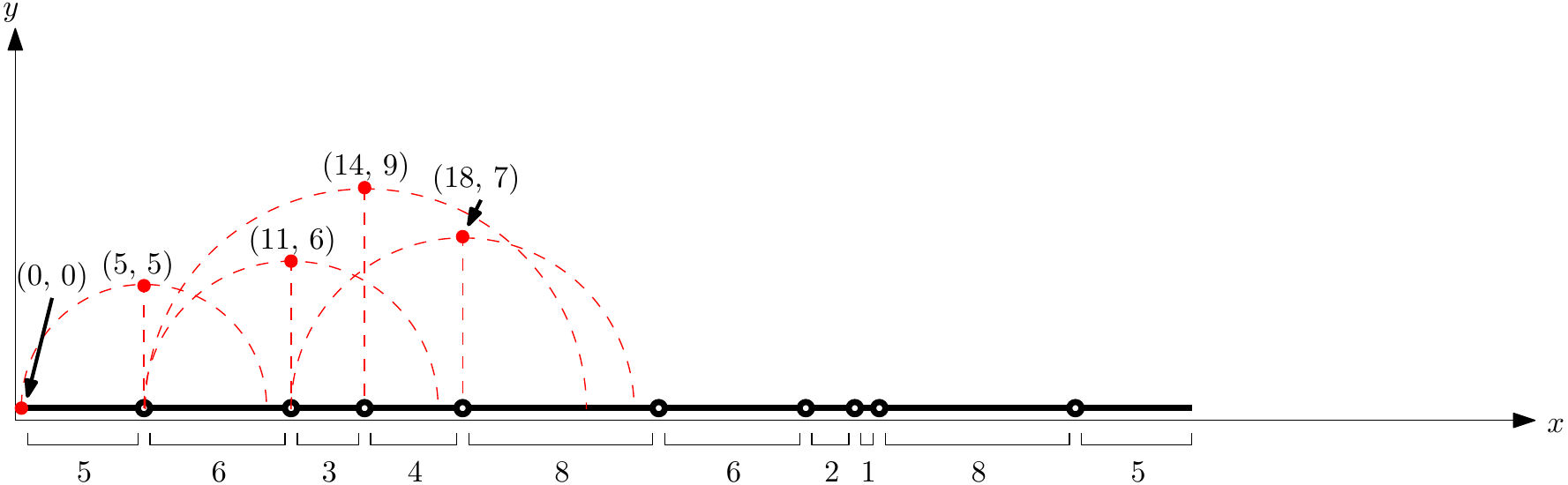}
\caption{Arcs for the first five hinges (including hinge 0) of the ruler with segments of lengths 5, 6, 3, 4, 8, 6, 2, 1, 8 and 5.}
\label{fig:partial}
\end{center}
\end{figure}

If we wrap the segments to the left of hinge $i$ into its wrapping length and then fold it, the wrapped segments trace an arc with apex $(x_i, y_i)$, where $x_i$ is the position of the hinge (the sum of the lengths of the segments to its left) and $y_i$ is its wrapping length.  Figure~\ref{fig:partial} shows the arcs for the first five hinges (including hinge 0) of the ruler with segments of lengths 5, 6, 3, 4, 8, 6, 2, 1, 8 and 5, with the red dots marking the apexes.  Notice that if we fold hinge 2 in Figure~\ref{fig:partial} then we cannot fold hinge 3, because $x_2 + y_2 = 17 > 14 = x_3$, and if we fold hinge 3 then we cannot fold hinge 4, because $x_3 + y_3 = 23 > 18 = x_4$.  When $x_h + y_h \leq x_i$, on the other hand, we can fold the segments to the left of hinge $h$ into its wrapping length $y_h$, fold hinge $h$, and then fold hinge $i$, so $y_i \leq x_i - x_h$.

\begin{observation}
\label{obs:wrapping_length}
For $i > 0$, the wrapping length $y_i$ of hinge $i$ is $x_i - x_h$, where hinge $h$ is the last previous hinge such that $x_h + y_h \leq x_i$.
\end{observation}

\begin{figure}[t]
\begin{center}
\includegraphics[width=\textwidth]{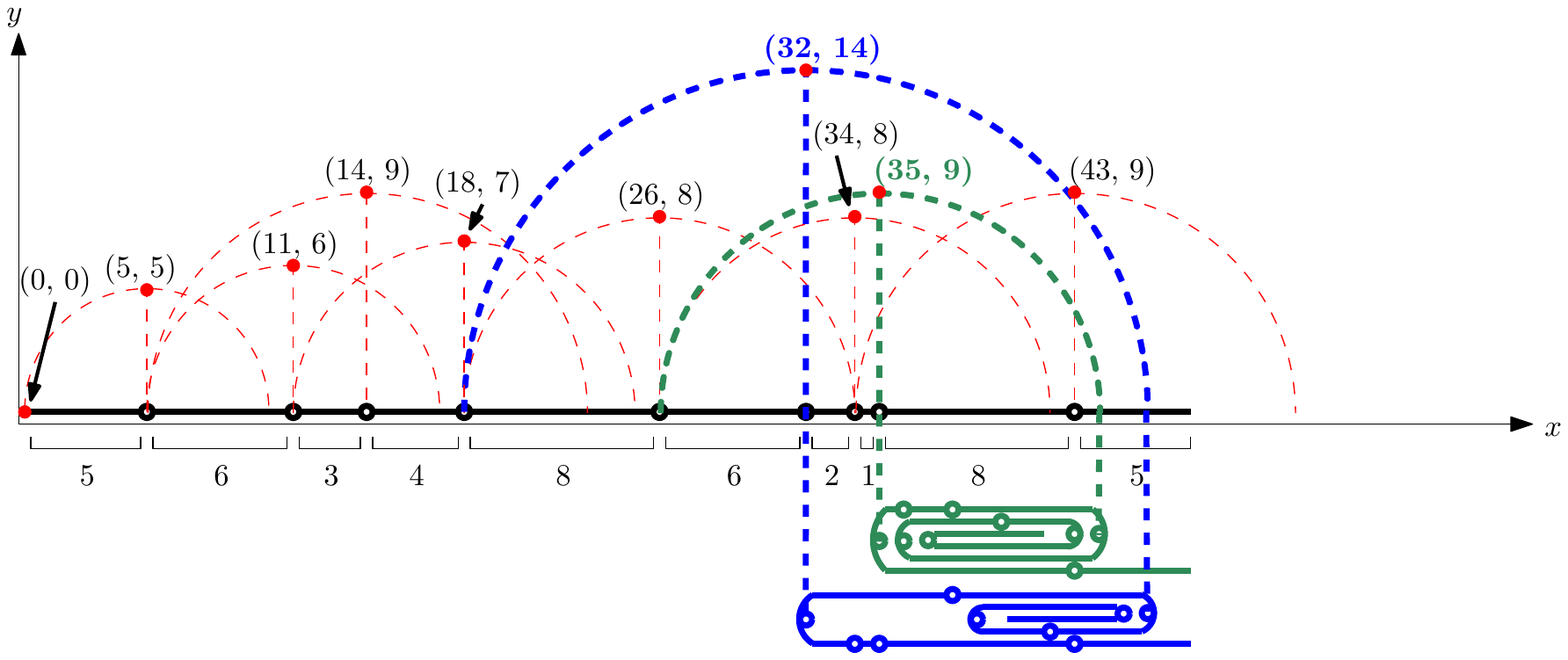}
\caption{A case in which choosing the last arc {\bf (blue)} that ends before or at hinge $n$ yields a larger wrapping length than choosing an arc {\bf (green)} that ends earlier.}
\label{fig:semi-final}
\end{center}
\end{figure}

Figure~\ref{fig:partial} suggests a simple quadratic-time dynamic program for computing the wrapping lengths: set $x_0 = 0$ and $y_0 = 0$ (because hinge 0 at the left end of the ruler has no segments to its left); for $i$ from 1 to $n$, set $x_i$ to the position of hinge $i$ and set $y_i$ to
\[x_i - \max \{x_h\ :\ h \leq i,\ x_h + y_h \leq x_i\}\]
It may seem at first that we can simply choose the last arc that ends before or at each hinge, but Figure~\ref{fig:semi-final} shows we are sometimes better off choosing an arc that ends earlier: the blue arc ends closer to the end of the ruler but the green arc has a later center and yields a smaller wrapping length.

It may also seem at first that the wrapping length $y_n$ of hinge $n$ should always be the smallest length into which we can wrap the whole ruler, but Figure~\ref{fig:final} shows this is guaranteed only when we require the distance from the last folded hinge to the end of the ruler to be at least the distance between the last two folded hinges.  (We display the wrappings as triangular spirals here to make it easier to show the segments' lengths.)  When we do not require this, we can scan the hinges in linear time to find the one that minimizes the maximum of its wrapping length and the distance to the end of the ruler; that maximum is the smallest length into which we can wrap the ruler when that hinge is the last one folded.  Summing up, so far we have the following result:

\begin{figure}[t]
\begin{center}
\includegraphics[width=\textwidth]{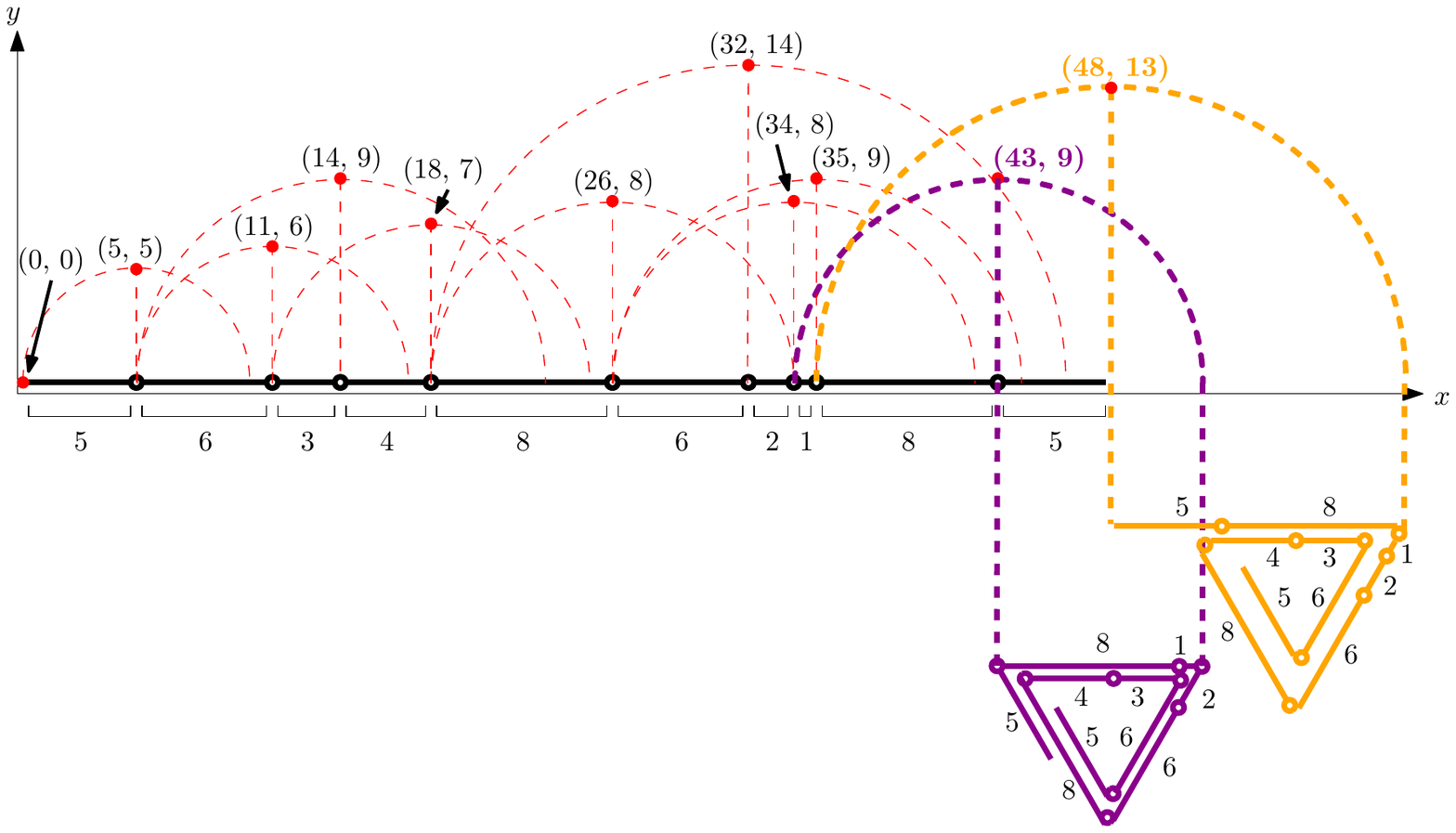}
\caption{A case in which the wrapping length of hinge $n$ {\bf (orange)} is more than the shortest length into which we can wrap the whole ruler {\bf (magenta)} when we do not require the distance from the last folded hinge to the end of the ruler to be at least the distance between the last two folded hinges.}
\label{fig:final}
\end{center}
\end{figure}

\begin{theorem}
\label{thm:quadratic}
Given the positive lengths of the $n$ segments of a carpenter's ruler, in $O (n^2)$ time we can compute the shortest length into which it can be wrapped.
\end{theorem}

\section{An $O (n \log n)$-time algorithm}
\label{sec:nlogn}

We can use a range-minimum data structure to reduce the running time in Theorem~\ref{thm:quadratic} to $O (n \log n)$, but this complicates the implementation somewhat.  Instead, recall the $O (n \log n)$-time array-based algorithm for finding a longest increasing subsequence of a list $L [1..n]$ of numbers, which Fredman~\cite{Fre75} analyzed and attributed to Knuth~\cite{Knu73}.  That algorithm starts with an array $T [1..n]$ with $T [1]$ set to $L [1]$ and the other entries empty; for $i > 1$, it compares $L [i]$ against the rightmost non-empty value $T [k]$ in $T$ and, if $L [i] > T [k]$, sets $T [k + 1] = L [i]$; otherwise, it performs a binary search in $T [1..k]$ --- which is always sorted --- to find the leftmost value $T [h] < L [i]$ and sets $T [h] = L [i]$.  By induction, this maintains the invariant that each $T [j]$ is always the smallest value such that a subsequence of $L [1..i]$ of length $j$ ends with $T [j]$.

A key idea behind Knuth's algorithm is that if we find $T [j] < L [i]$ then we need not keep the current value of $T [j]$ because any extension of a subsequence ending with $T [j]$ is also an extension of a subsequence ending with $L [i]$.  We can apply a similar idea to obtain an $O (n \log n)$-time array-based algorithm for ruler wrapping: we start with an array $P [0..n]$ of pairs with $P [0] = (x_0, y_0) = (0, 0)$ and the other entries empty; for $i > 1$, we
\begin{enumerate}
\item add the length of the $i$th segment of the ruler to $x_{i - 1}$ to obtain $x_i$,
\item perform a binary search in the non-empty prefix $P [0..k]$ of $P$ --- which is always sorted both by $x$-coordinate and by sum $x + y$ --- to find the rightmost pair $(x_h, y_h)$ with $x_h + y_h \leq x_i$ (there always is such a pair, since $x_0 + y_0 = 0$),
\item set $y_i = x_i - x_h$,
\item scan leftward from $P [k]$ discarding pairs $(x_j, y_j)$ with $x_j + y_j \geq x_i + y_i$,
\item insert $(x_i, y_i)$ immediately to the right of the rightmost undiscarded pair.
\end{enumerate}

To see why $P [0..k]$ is always sorted both by $x$-coordinate and by sum $x + y$, suppose it is sorted before we process the length of the $i$th segment of the ruler, and consider that $x_i$ is larger than any previous $x$-coordinate and we discard all the pairs $(x_j, y_j)$ with $x_j + y_j \geq x_i + y_i$.  To see why we can discard any such pair $(x_j, y_j)$, consider that we will never choose the arc centered at $x_j < x_i$ that ends at $x_j + y_j \geq x_i + y_i$, when we can choose the arc centered at $x_i$.  Finally, to see why processing the length of the $i$th segment of the ruler takes us $O (\log n)$ amortized time, consider that the binary search takes $O (\log n)$ time, we can stop discarding pairs as soon as we encounter one that sums to less than $x_i + y_i$, and we can charge each pair we discard to the segment of the ruler for which we inserted it.

We wrote earlier that when we do not require the distance from the last folded hinge to the end of the ruler to be at least the distance between the last two folded hinges, we can scan the hinges in linear time to find the one that minimizes the maximum of its wrapping length and the distance to the end of the ruler.  We can no longer scan all of the hinges easily if we discard some pairs, but we cannot discard the pair for what should be the last folded hinge.  To see why, consider that our algorithm works online and let hinges $g$ and $h$ be the last folded hinges in a shortest wrapping; if a segment of length $(x_h - x_h) - (x_n - x_h)$ were appended to the ruler, then we would need to have $(x_h, y_h)$ in the array in order to compute $y_{n + 1} = x_{n + 1} - x_h$.  If a segment of length 3 were appended to the ruler in Figure~\ref{fig:final}, for example, then because $x_9 + y_9 = 43 + 9 \leq x_{11} = 51$, we would have $y_{11} = x_{11} - x_9 = 8$ --- so our algorithm cannot discard $(x_9, y_9)$.

\begin{theorem}
\label{thm:nlogn}
Given the positive lengths of the $n$ segments of a carpenter's ruler, in $O (n \log n)$ time we can compute the shortest length into which it can be wrapped.
\end{theorem}

\section{A linear algorithm}
\label{sec:linear}

Fredman showed the number of comparisons Knuth's algorithm performs is the best possible for finding a longest increasing subsequence, to within a linear term.  Rather surprisingly, however, we can further reduce the running time in Theorem~\ref{thm:nlogn} to $O (n)$.  To see why, consider that for ruler wrapping, because the segments' lengths are positive, the $x_i$s are themselves an increasing sequence.  Therefore, if we are currently processing $x_i$ and we have $(x_g, y_g)$ and $(x_h, y_h)$ in our array with $x_g < x_h < x_i$ and $x_h + y_h \leq x_i$, then not only will we not choose the arc centered on $x_g$ to compute $y_i$, we will never choose it to compute any other $y$-coordinate in the future, either.  It follows that instead of using binary search to find the rightmost pair $(x_h, y_h)$ in our array with $x_h + y_h \leq x_i$, we can scan rightward from the first non-empty cell in our array, discarding pairs until we find that rightmost pair $(x_h, y_h)$ with $x_h + y_h \leq x_i$.  Again, we charge each pair we discard to the segment of the ruler for which we inserted it.  Table~\ref{tab:trace} shows the contents of our array while we process our running example.  We note that the rightmost cell $P [10]$ is always empty in this example, but its presence guarantees we have space for all the pairs even if we never discard pairs on the right.

\begin{table}[t]
\begin{center}
\caption{The contents of our array $P$ while processing our running example.}
\smallskip
\label{tab:trace}
\resizebox{\textwidth}{!}
{\begin{tabular}{r|ccccccccccc}
step & $P [0]$ & $P [1]$ & $P [2]$ & $P [3]$ & $P [4]$ & $P [5]$ & $P [6]$ & $P [7]$ & $P [8]$ & $P [9]$ & $P [10]$ \\
\hline\\[-1ex]
 0 & $(0, 0)$ & & & & & & & & & & \\
 1 & $(0, 0)$ & $(5, 5)$ & & & & & & & & & \\
 2 &          & $(5, 5)$ & $(11, 6)$ & & & & & & & & \\
 3 &          & $(5, 5)$ & $(11, 6)$ & $(14, 9)$ & & & & & & & \\
 4 &          &          & $(11, 6)$ & $(14, 9)$ & $(18, 7)$ & & & & & & \\
 5 &          &          &           &           & $(18, 7)$ & $(26, 8)$ & & & & & \\
 6 &          &          &           &           & $(18, 7)$ & $(26, 8)$ & $(32, 14)$ & & & & \\
 7 &          &          &           &           &           & $(26, 8)$ & $(34, 8)$  & & & & \\
 8 &          &          &           &           &           & $(26, 8)$ & $(34, 8)$  & $(35, 9)$ & & & \\
 9 &          &          &           &           &           &           & $(34, 8)$  & $(35, 9)$ & $(43, 9)$ & & \\
10 &          &          &           &           &           &           &            & $(35, 9)$ & $(43, 9)$ & $(48, 13)$ &
\end{tabular}}
\end{center}
\end{table}

Even when we do not require the distance from the last folded hinge to the end of the ruler to be at least the distance between the last two folded hinges, during our rightward scan we cannot accidentally discard a pair we might need later, this time because we discard only pairs $(x_g, y_g)$ with $x_g + y_g \leq x_i$ --- so the distance from hinge $g$ to the end of the ruler is at least hinge $g$'s wrapping length, which is the distance from the previous folded hinge if $g$ is the last folded one.

\begin{theorem}
\label{thm:linear}
Given the positive lengths of the $n$ segments of a carpenter's ruler, in $O (n)$ time we can compute the shortest length into which it can be wrapped.
\end{theorem}

\section{Maximizing folded hinges}
\label{sec:maximizing}

If we require the distance from the last folded hinge to the end of the ruler to be at least the distance between the last two folded hinges (considering hinge 0 to be folded), then we can wrap a ruler into length $\ell$ if and only if we can partition the string of the segments' length into substrings such that the totals of the substrings are non-decreasing and the last substring's total is at most $\ell$.

It is also fairly natural to ask how quickly we can partition the string into the maximum number of substrings whose totals are non-decreasing (in other words, to wrap a given carpenter's ruler while maximizing the number of folded hinges).  Notice there are ways to maximize the number of folded hinges that do not minimize the length (such as wrapping 2, 1, 3 into length 4 with one folded hinge, between the segments of lengths 2 and 1), and ways to minimize the length that do not maximize the number of folded hinges (such as wrapping 1, 1, 3 into length 3 with one folded hinge, between the segments of lengths 1 and 3).  Nevertheless, our algorithm also always maximizes the number of folded hinges.

To see why, assume our claim does not hold and consider a shortest sequence of lengths $L [1..n]$ for which it fails.  Suppose our algorithm partitions $L$ into $k$ substrings $L_1, \ldots, L_k$ with non-decreasing totals, but it is possible to partition $L$ into $k' > k$ substrings $L_1', \ldots, L_{k'}'$ with non-decreasing totals.  Notice $L_1'$ contains $L_1 [1]$; by the correctness of our algorithm for ruler wrapping, the total of $L_{k'}'$ is at least the total of $L_k$ and so $L_{k'}'$ contains $L_k [1]$.  Since $k < k'$, it follows from the pigeonhole principle that there is some $j$ with $2 \leq j \leq k' - 1$ such that $L_{j - 1}'$ contains $L_{j - 1} [1]$ and $L_{j + 1}'$ contains $L_j [1]$.

The concatenation of $L_1, \ldots, L_{j - 1}$ can be partitioned into $L_1', \ldots, L_{j - 1}', L_j''$, with $L_j''$ containing all the lengths in $L_j'$ (and perhaps some from $L_{j + 1}'$) and thus totalling at least $t_j' \geq t_{j - 1}'$.  In other words, this concatenation can be partitioned into $j$ substrings with non-decreasing totals, whereas our algorithm partitions it into $j - 1$ substrings, contradicting our choice of $L$ as short as possible.

\begin{corollary}
\label{cor:linear}
Given a sequence of $n$ positive numbers, in $O (n)$ time we can partition it into the maximum number of substrings whose totals are non-decreasing.
\end{corollary}

\section{Experiments}
\label{sec:experiments}

Another natural question is how quickly the minimum length into which we can wrap a ruler tends to increase as the number of segments increases, assuming the segments' lengths are drawn from a uniform distribution.  It may be that distributions of these minimum lengths have some connection to distributions of lengths of longest increasing subsequences, which have been extensively studied.

To begin the study of these wrapping distributions, we implemented our algorithm in {\tt C} (the key subroutine is shown in Figure~\ref{fig:code}), and ran it on 1000 pseudo-randomly generated lists of 100000 integers from 1 to 100.  For each $i$ from 0 to 10000 we computed the average wrapping length of the $i$th hinge over the runs.  The averages for hinges 0, 1, 10, 100, 1000, 10000 and 100000 were 0, 50.9, 188.0, 547.6, 1624.2, 4948.4 and 15323.4 --- so it seems the wrapping length grows roughly as $\sqrt{n}$ (times the midpoint of the range from which the integers are drawn).  It might also be interesting for what sequence of $n$ numbers from a bounded range the wrapping length of hinge $n$ is maximum; it is not $n$ copies of the largest number in the range, since a ruler of any number of segments of length 100, for example, can be wrapped into length 100.

\begin{figure}[t!]
\begin{quotation}
\noindent {\tt \begin{verbatim}
struct pair {
  int x;
  int y;
};
\end{verbatim}
$\vdots$
\begin{verbatim}
int wrap(int *L, int n) {
  
  struct pair P[n + 1];
  P[0].x = P[0].y = head = tail = 0;
  
  for (int i = 1; i <= n; i++) {
    int x = P[tail].x + L[i - 1];
    
    while (head != tail &&
      P[head + 1].x + P[head + 1].y <= x) {
      head++;
    }
    
    int y = x - P[head].x;
    
    while (P[tail].x + P[tail].y >= x + y) {
      tail--;
    }
    
    tail++;
    P[tail].x = x;
    P[tail].y = y;
  }
  
  return(P[tail].y);
}
\end{verbatim}}
\end{quotation}
\caption{The key subroutine for computing the smallest length into which we can wrap a ruler whose segments' lengths are given in an array {\tt L} of {\tt n} {\tt ints}.}
\label{fig:code}
\end{figure}

If we implement $P$ as a linked list instead of as an array and delete nodes when we discard the pairs they store (or implement it as an resizable circular array), then our algorithm's workspace is proportional to the maximum number of pairs stored in $P$ at any point during the execution.  Over the 1000 runs of our program, we checked the maximum number of pairs stored when processing each hinge, and those maxima seem to grow roughly as $\sqrt{n}$ as well; this means our algorithm is online but not properly streaming.  In contrast, at least when we require the distance from the last folded hinge to the end of the ruler to be at least the distance between the last two folded hinges, the obvious greedy algorithm for ruler wrapping --- fold whenever possible --- needs to keep track only of the distance from the most recent folded hinge and the distance between that and the folded hinge before it, so it uses linear time and constant workspace and is streaming.  We implemented this algorithm as well and in practice it seems to have an approximation ratio at most about $5 / 4$, even though its worst-case ratio is $\Omega (\sqrt{n})$ (consider a ruler whose segments have lengths $2, 1, 3, 3, 2, 1, 3, 3, 3, 3, 2, 1, 3, 3, 3, 3, 3, 3, 2, 1, \ldots$, with $2 i$ copies of 3 between the $i$th and $(i + 1)$st copies of 2 and 1).  We are curious whether there is a streaming algorithm with a reasonable worst-case ratio.

\section{Acknowledgments}
\label{sec:acknowledgments}

Many thanks to Joseph O'Rourke for proposing ruler wrapping, and to Joseph Mitchell, David Wagner and the other participants of the CCCG '21 open-problem session.  The first author thanks NSERC for his Discovery Grant, RGPIN-2020-07185, which funded this research; his mother, Meg Gagie, for proofreading; and his CSCI 3110 class, for trying so desperately to avoid building range-minimum data structures that they implemented Knuth's algorithm instead --- and thus inadvertently taught it to him.


\begin{thebibliography}{1}

\bibitem{CD05}
Gruia C{\u{a}}linescu and Adrian Dumitrescu.
\newblock The carpenter’s ruler folding problem.
\newblock {\em Combinatorial and Computational Geometry}, 52:155, 2005.

\bibitem{Fre75}
Michael~L. Fredman.
\newblock On computing the length of longest increasing subsequences.
\newblock {\em Discrete Mathematics}, 11(1):29--35, 1975.

\bibitem{HJW85}
John Hopcroft, Deborah Joseph, and Sue Whitesides.
\newblock On the movement of robot arms in 2-dimensional bounded regions.
\newblock {\em SIAM Journal on Computing}, 14(2):315--333, 1985.

\bibitem{Knu73}
Donald~E. Knuth.
\newblock {\em The Art of Computer Programming}, volume~3.
\newblock Addison-Wesley, 1973.

\end{thebibliography}
\end{document}